# Recent Technological Developments on LGAD and iLGAD Detectors for Tracking and Timing Applications

G. Pellegrini[1], M. Baselga[1], M. Carulla[1], V. Fadeyev[2], P. Fernández-Martínez[1], M. Fernández García[4], D. Flores[1], Z. Galloway[2], C. Gallrapp[3], S. Hidalgo[1], Z. Liang[2], A. Merlos[1], M. Moll[3], D. Quirion[1], H. Sadrozinski[2], M. Stricker[3], I.Vila[4]

1) Centro Nacional de Microelectrónica, IMB-CNM-CSIC, Barcelona, Spain
2) Santa Cruz Institute of Particle Physics SCIPP, Santa Cruz, CA, USA
3) CERN, Geneva, Switzerland
4) Instituto de Física de Cantabria IFCA-CSIC-UC, Santander, Spain

**Abstract**

This paper reports the last technological development on the Low Gain Avalanche Detector (LGAD) and introduces a new architecture of these detectors called inverse-LGAD (iLGAD). Both approaches are based on the standard Avalanche Photo Diodes (APD) concept, commonly used in optical and X-ray detection applications, including an internal multiplication of the charge generated by radiation. The multiplication is inherent to the basic $n^{++}$-$p^+$-$p$ structure, where the doping profile of the $p^+$ layer is optimized to achieve high field and high impact ionization at the junction.

The LGAD structures are optimized for applications such as tracking or timing detectors for high energy physics experiments or medical applications where time resolution lower than 30 ps is required. Detailed TCAD device simulations together with the electrical and charge collection measurements are presented through this work.

**Keywords**: silicon detectors, avalanche multiplication, timing detectors, tracking detectors.

## 1. Introduction

All avalanche diode detectors [1] have a region with a high electrical field so as to cause multiplication of signal charges (electron and/or holes) flowing through this region. The gain mechanism is achieved within the semiconductor material by raising the electric field as high as necessary to enable the drifting electrons to create secondary ionization during the collection process. Normally, the junction is made up of a thin and highly doped n-type layer on top of a moderately doped p-layer in which the multiplication (of electrons) takes place. A high resistivity p-type silicon substrate is typically used to fabricate detectors with a bulk that can be fully depleted.

Compared to standard APD detectors, LGAD (Low Gain Avalanche Detectors) and iLGAD (inverse LGAD) structures analyzed in this work exhibit moderate gain values that are mandatory to obtain fine segmentation pitches in the fabrication of microstrip and pixel detectors free from the limitations commonly found in avalanche detectors [2]. In addition, a moderate multiplication allows the fabrication of thinner detectors with the same output signal of detectors without internal gain integrated on standard thick substrates. The design of LGAD and iLGAD structures exploits the charge multiplication effect to obtain a silicon detector that can concurrently measure with high accuracy time and space at once or separately.

The process technology of LGAD detectors has already been introduced in [3] while radiation hardness studies are reported in [4]. Several wafers were fabricated with a highly doped p-type layer, implanted under the $n^+$ shallow diffusion to reach the electric field strength [5] that provides the required gain. The structure is similar to the one of the Avalanche Photo Diodes but optimized for operation in the linear gain region. With proper electrode isolation techniques the production of segmented sensors with breakdown voltage in excess of 1000 V is feasible.

The aim of this work is to describe the operation mode and performance of LGAD and iLGAD sensors. The internal gain in silicon sensors and its uniformity in segmented detectors are discussed. A new approach to the charge collection is also analyzed for the iLGAD structures where the segmentation is implemented at the ohmic contact ($p^+$ diffusion at the back side of the LGAD), thus ensuring a uniform multiplication at the $n^+p$ junction side. I(V) and C(V) performances and charge collection data obtained from α and laser measurements to get an estimation of the gain, are included. Finally, preliminary TCAD device simulation results of iLGAD detectors are reported.

2. **Low Gain Avalanche detectors (LGAD)**

Pad LGAD detectors have been fabricated and extensively measured. Three different designs were considered for the edge termination, based on the location of a deep n-type diffusion called Junction Termination Extension (JTE) [6], but all the detectors include a shallow n-type collector ring between the edge of the main junction and the device periphery to collect the undesired leakage current generated at the periphery region. Design 1 includes a JTE diffusion both in the main junction and the collector ring while design 2 the JTE is only placed at the collector ring. Finally, design 3 does not include the JTE diffusion but only the shallow n+ implant. A detailed study of the termination structures used for the LGAD detectors is reported in [7].

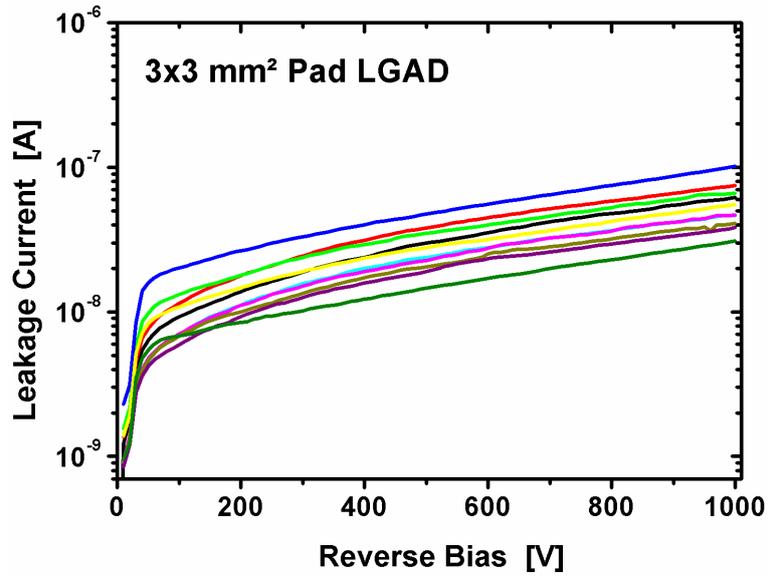

**Fig. 1.** I(V) curves of 300 μm pad LGAD detectors with edge termination according to design 1 and detection area of 3x3 mm².

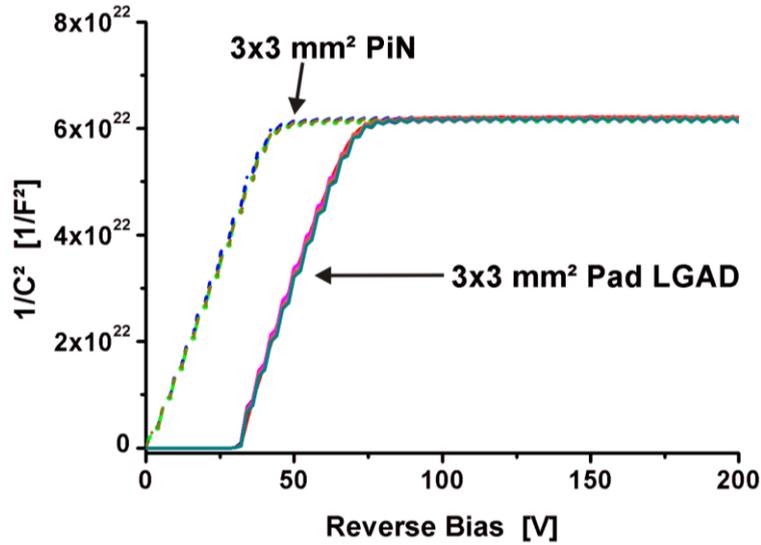

**Fig. 2.** $1/C^2$(V) curves of 300 μm pad LGAD and PiN detectors with edge termination according to designs 1, 2 and 3 and detection area of 3x3 mm².

LGAD detectors with the described edge termination designs were fabricated at the IMB-CNM Clean Room on 300 μm substrates. The last processed wafers have been extensively measured to check the electrical performances, the charge collection capability and the yield. The I(V) curves plotted in Fig. 1 correspond to different detectors with edge termination according to design 1, the one with the highest voltage capability, and the optimum implantation dose of the p-well to obtain a gain of 10. It can be clearly envisaged that the leakage current values in the multiplication area (3x3 mm²) are in the range of 10-30 nA at a reverse bias of 500 V with a yield in the range of 80%. The $1/C^2$ (V) curves plotted in Fig. 2 show the different depletion spread in the pad LGAD detectors when compared with the PiN detectors fabricated within the same process technology only skipping the p-well implantation step. From these figures the reverse voltage at which the p-well becomes completely depleted (30 V) and the full depletion voltage (80 V) can be inferred.

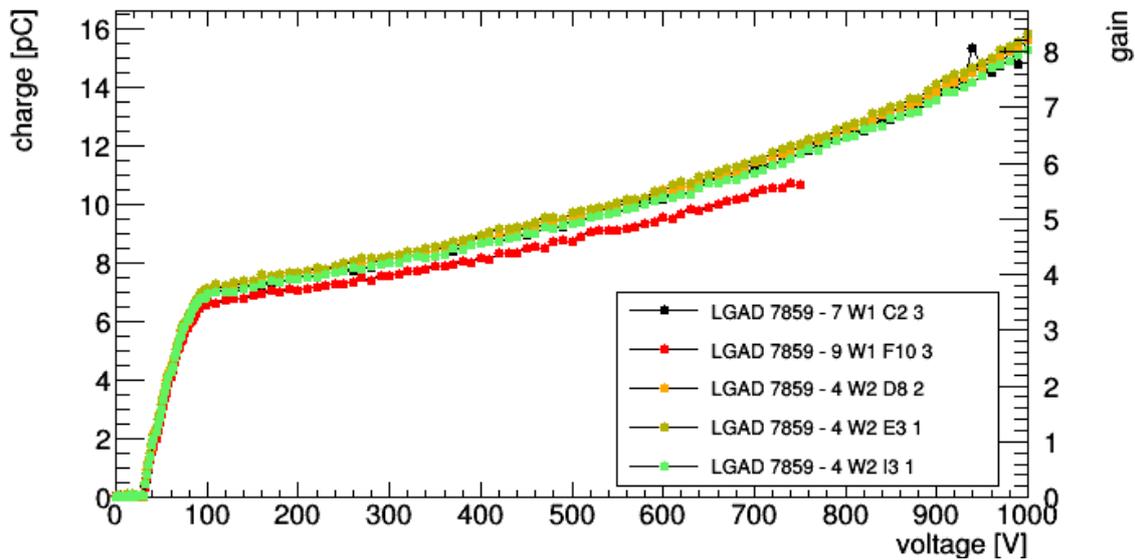

**Fig. 3.** Experimental collected charge from IR laser vs. applied reverse voltage for different LGAD detectors with optimum P-well implantation dose.

LGAD samples were characterized at IMB-CNM, Santa Cruz Institute of Particle Physics (SCIPP) and CERN. The gain of different samples with P-well implantation dose is shown in Fig.3, where the good linearity in the typical operating voltages (200 to 800 V) can be observed. The gain in this voltage range varies from 4.1 to 6.6. Measurements with backside illumination, using an infrared laser, revealed that a high uniformity of the electric field in the detection region is achieved.

Pad LGAD detectors can only provide a signal proportional to the energy deposited by the incident particle. On the contrary, strip LGAD detectors provide combined information on energy deposited and position of the incident particle and are crucial for high accuracy tracking applications. LGAD detector with strip electrode pattern were also fabricated within the same wafer batch. Preliminary results show multiplication at the central part of each strip, as shown in Fig. 4 where the charge collected scanning three LGAD strips with a red laser from the front side is plotted for different reverse bias values. The top surface of the strips illuminated by the laser is not covered with Aluminum. At a reverse bias of 50 V (Fig. 4 left) the substrate is not yet depleted and no multiplication is present in the structure. On the contrary, when the reverse bias is increased to 150 V (Fig. 4 right) the substrate is fully depleted and the multiplication is already active at the central part of each strip. Using the signal of 20mV obtained with a reference PiN diode (no multiplication), the experimental gain value for the strip sensors at 150 V is around 2. The experimental evolution of the charge collection vs. the reverse voltage in strip LGAD detectors is provided in Fig. 5, where a gain of 2 is obtained at 200 V.

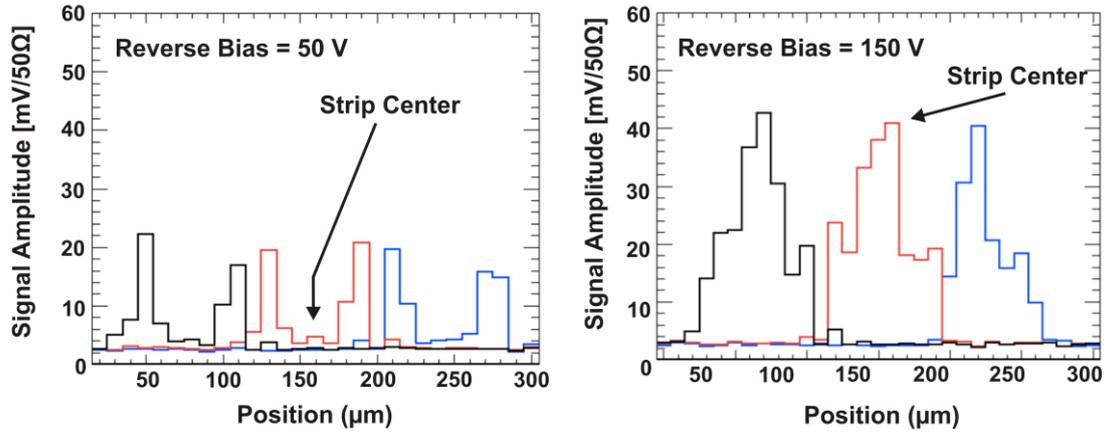

**Fig. 4.** Red laser scanning of three inner LGAD strips before full depletion (left) and after full depletion (right).

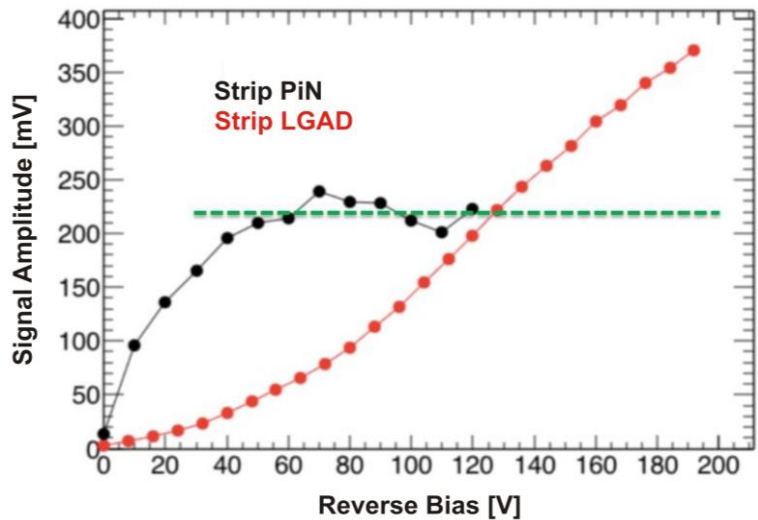

**Fig. 5.** Preliminary experimental results of charge collection measurements on strip LGAD and strip PiN detectors vs. the applied reverse voltage. The laser was illuminating the middle of the strip.

### 3. Inverse Low Gain Avalanche detectors (iLGAD)

Although strip LGAD detectors with low gain values were successfully fabricated, spatial non-uniformities due to the different multiplication from center to edge of each strip were also measured (see Fig. 4). In this sense, the strip iLGAD structure drawn in Fig. 6 is designed to maximize the multiplication area while the segmentation is transferred to the $p^+$ ohmic side. In this sense, a strip p-on-p detector [8] with internal gain is implemented where the collected signal is due to holes flowing back from multiplication junction. As a consequence, the detection time is increased in comparison with the LGAD counterparts due to the lower mobility of holes. There are two possible ways to reduce the detection time: using thinner substrates with the inherent handling difficulty during their clean room processing or implementing the iLGAD in the active silicon part of a Silicon-on-Insulator (SOI) wafer and then removing the handling substrate and the buried oxide.

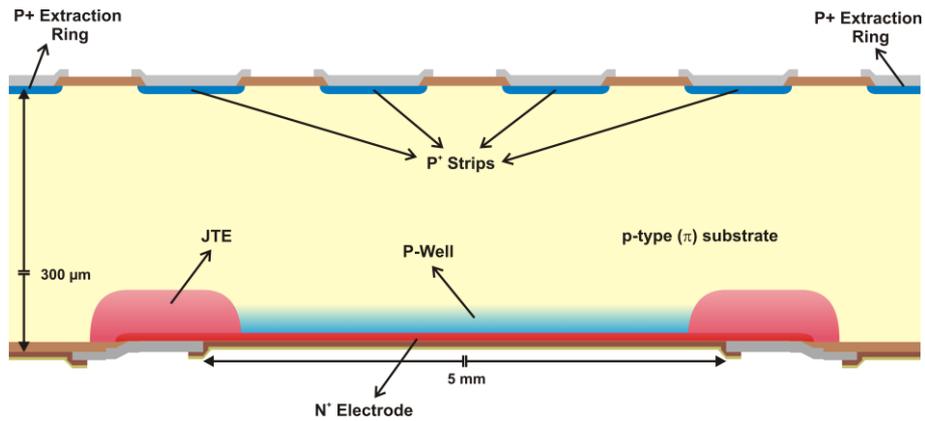

**Fig. 6.** Cross-section of the proposed strip iLGAD structure.

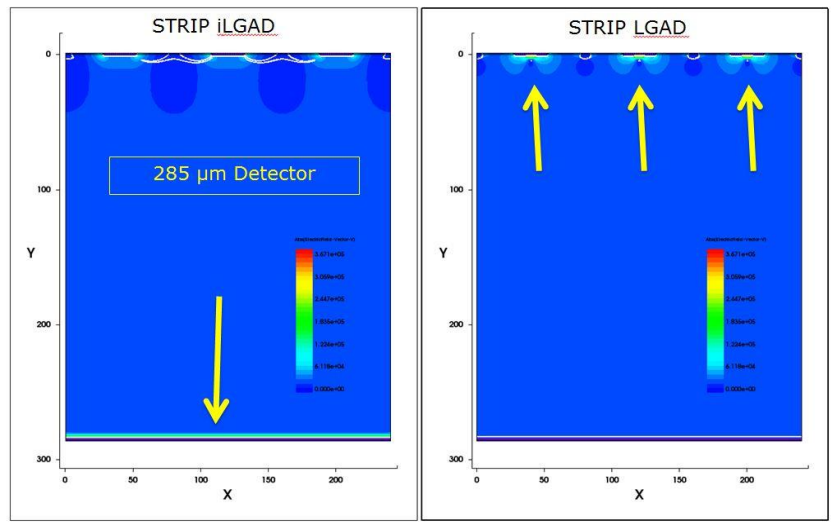

**Fig. 7.** Electric field distribution in the detection area of equivalent strip iLGAD (left) and strip LGAD (right). The arrows show the high electric field area in the two structures.

TCAD simulations with the Sentaurus software [9] have indicated that the voltage capability of the strip iLGAD is similar to that of the equivalent strip LGAD (> 1000 V) although the maximum electric field moves from the segmentation side, as it is the case in strip LGADs, to the multiplication side. A slight increase of the leakage current is observed in the strip iLGAD due to the higher multiplication area in comparison with the strip LGAD where the multiplication side is segmented. The electric field distribution at high reverse bias has been also simulated for both structures and it is plotted in Fig. 7. The strip LGAD structure exhibits a narrow high electric field region at the central part of each strip while the region of maximum electric field is uniformly distributes at the multiplication side in the case of the strip iLGAD.

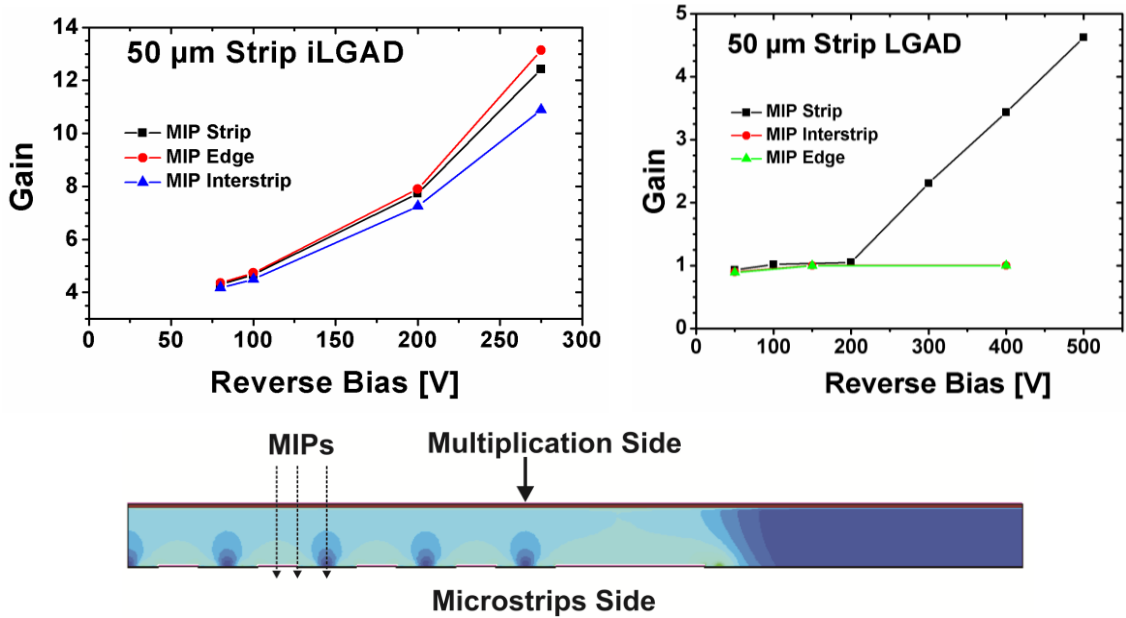

**Fig. 8.** Gain simulation by using MIP on strip iLGAD (left) and strip LGAD (right) structures. The position of the incident particles in the strip iLGAD is drawn at the bottom of the figure.

The charge collection efficiency of the proposed iLGAD structure has been simulated by using minimum ionizing particles (MIP) and compared with the strip LGAD counterpart. The gain simulation results reported in Fig. 8 (left) clearly show that the strip iLGAD exhibits multiplication wherever the MIP enters the structure since the multiplication junction is no longer segmented. In this particular case, the iLGAD is implemented on 50 μm silicon layer of a SOI substrate that becomes fully depleted at 70 V. On the contrary, strip LGAD detectors only have gain when the incident particle crosses the central part of a strip, as shown in Fig. 8 (right). The latter is simulated on 285 μm thick substrates that become fully depleted at 180 V, where the multiplication starts to be relevant.

Strip iLGAD detectors are now being fabricated at IMB-CNM clean room including different design options and process technology variations to determine the optimum parameters to get a stable gain along the operating reverse voltage range and accurate detection performances. Strip iLGAD detectors are initially fabricated on conventional p-type silicon substrates but SOI substrates will also be used in next batches to obtain thin detectors with fast timing detection capabilities [10].

## 4. Conclusions

The electrical performances of the fabricated strip LGAD detectors demonstrate a voltage capability higher than 1000 V with leakage currents in the 20 nA/cm$^2$ range and a linear gain in the typical operating reverse voltage values (200 to 800 V) in the range of 5-10. However, red laser scanning measurements revealed a non-uniform multiplication across the strips, basically due to technological constrains.

The proposed strip iLGAD structure provides similar voltage capability and leakage current with a uniform gain across the strips since the multiplication junction is not segmented, as derived from MIPs simulation results.

**Acknowledgments**

This work was developed in the framework of the CERN RD50 collaboration and financed by the Spanish Ministry of Economy and Competitiveness through the Particle Physics National Program (FPA2013-48308-C2-2-P, FPA2013-48387-C6-2-P and FPA2013-48387-C6-1-P). This project has received funding from the European Union's Horizon 2020 Research and Innovation programme under Grant Agreement no. 654168 (AIDA-2020).